\documentclass[journal=jacsat,manuscript=article]{achemso}

\usepackage{setspace}
\usepackage{xcolor}
\usepackage{graphicx}
\usepackage{academicons}
\usepackage{orcidlink}
\usepackage{multirow}
\usepackage{chemformula}
\usepackage{makecell}
\usepackage{colortbl}
\usepackage{enumitem}
\usepackage{placeins}
\usepackage{pifont}
\usepackage{hyperref}
\usepackage{xr-hyper}
\author{\normalsize Zhuohan Li\,\orcidlink{0000-0001-5372-9450}}
\affiliation{\normalsize Materials Sciences Division, Lawrence Berkeley National Laboratory, Berkeley, California 94720, United States}

\author{\normalsize Benjamin X. Lam\,\orcidlink{0000-0002-2956-7719}}
\affiliation{\normalsize Department of Materials Science and Engineering, University of California, Berkeley, California 94720, United States}

\author{\normalsize Shilong Wang\,\orcidlink{0009-0004-8504-5802}}
\affiliation{\normalsize Department of Materials Science and Engineering, University of California, Berkeley, California 94720, United States}

\author{\normalsize Gerbrand Ceder \orcidlink{0000-0001-9275-3605}}
\email{Correspondence: gceder@berkeley.edu}
\affiliation{\normalsize Materials Sciences Division, Lawrence Berkeley National Laboratory, Berkeley, California 94720, United States}
\alsoaffiliation{\normalsize Department of Materials Science and Engineering, University of California, Berkeley, California 94720, United States}

\date{\today}

\title{\ch{Li+}/\ch{H+} exchange in solid-state oxide Li-ion conductors}

\abbreviations{IR,NMR,UV}

\begin{document}

\setstretch{2}

\begin{abstract}
\noindent\textbf{Abstract:} 
Understanding the moisture stability of oxide Li-ion conductors is important for their practical applications in solid-state batteries. Unlike sulfide or halide conductors, oxide conductors generally better resist degradation when in contact with water, but can still undergo topotactic \ch{Li+}/\ch{H+} exchange (LHX). Here, we combine density functional theory (DFT) calculations with a machine-learning interatomic potential model to investigate the thermodynamic driving force of the LHX reaction for two representative oxide Li-ion conductor families: garnets and NASICONs. Li-stuffed garnets exhibit a strong driving force for proton exchange due to their high Li chemical potential. In contrast, NASICONs demonstrate a higher resistance against proton exchange due to the lower Li chemical potential and the lower O-H bond covalency for polyanion-bonded oxygens. Our findings reveal a critical trade-off: Li stuffing enhances conductivity but increases moisture susceptibility. This study underscores the importance of designing Li-ion conductors that possess both high conductivity and high stability in practical environments.
\end{abstract}


\maketitle

\setstretch{2}

\begin{figure*}[t]
\centering
\includegraphics[width=\linewidth]{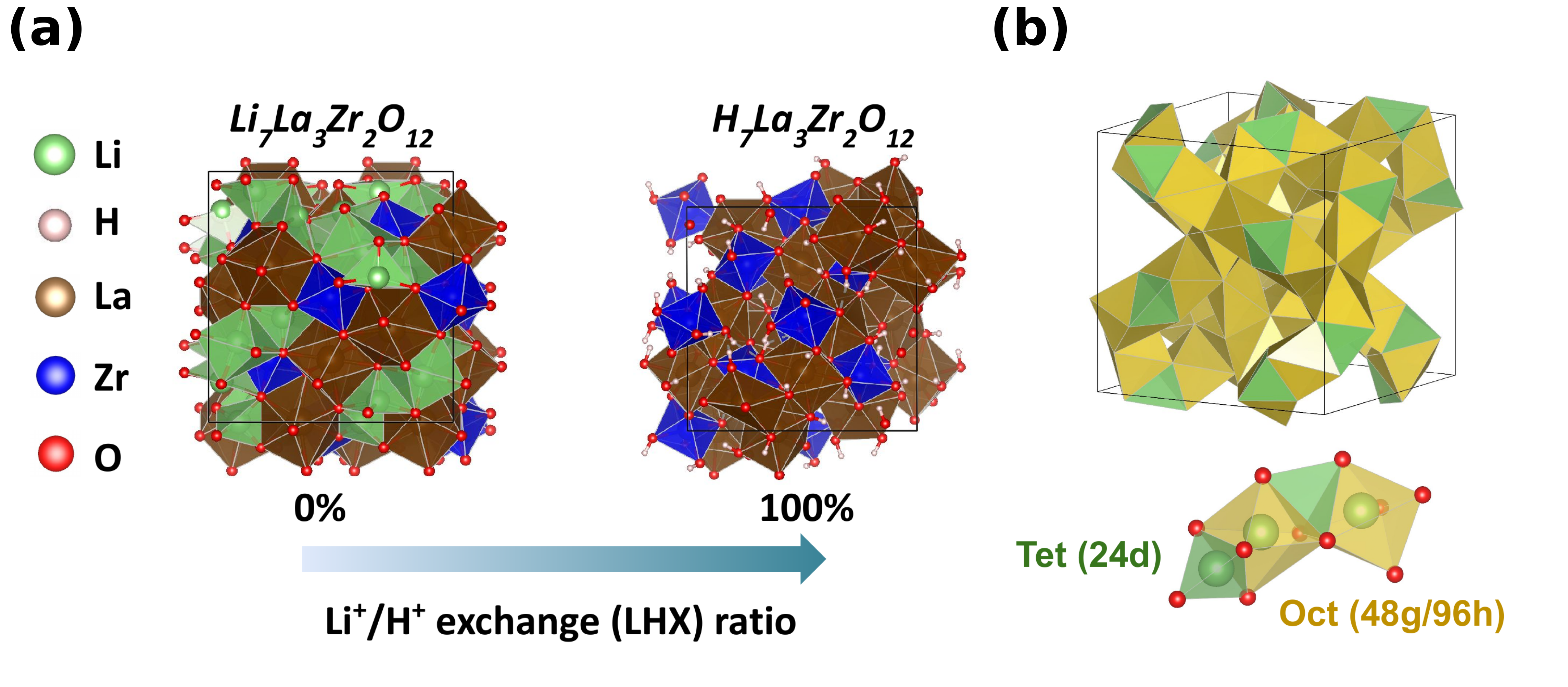}
\caption{(a) Schematic illustration of \ch{Li+}/\ch{H+} exchange (LHX) reaction in \ch{Li7La3Zr2O12}, (b) Li sites in garnet framework percolates through the interconnected tetrahedral (24d) and octahedral (48g/96h) sites. }
\label{fig:LHX_schematic} 
\end{figure*}

Solid-state batteries are considered as a replacement of the current commercial liquid electrolyte-based Li-ion batteries. The development of solid-state electrolytes (SSEs) with high Li-ion conductivity and stability in various environments is critical for the practical application of this technology. Among various types of SSEs, oxide-based electrolytes stand out due to their superior chemical stability in ambient air. In contrast to sulfide and halide SSEs, which are highly moisture-sensitive and decompose to release gases such as \ch{H2S} or \ch{HCl}, oxide SSEs generally maintain better chemical and structural integrity. However, some oxide SSEs, especially garnets, readily react with water through a topotactic ion exchange between Li ions in the bulk and protons from water \cite{Ye2021_LHX_review}. In Figure \ref{fig:LHX_schematic} (a), we schematically illustrate this \ch{Li+}/\ch{H+} exchange (LHX) reaction for a garnet compound \ch{Li7La3Zr2O12}, where all \ch{Li+} is exchanged with \ch{H+}, resulting in a fully protonated phase \ch{H7La3Zr2O12}. The introduced protons preferentially form O–H bonds with \ch{O^2-} anions in the polyhedra originally occupied by Li ions \cite{Ishii2024_H_diffusion}, while the host crystal structure consisting of edge-sharing \ch{LaO8^{15-}} and \ch{ZrO6^{8-}} polyhedra remains intact. When a garnet SSE is exposed to ambient air, the LHX reaction results in the formation of a protonated surface layer along with the accumulation of undesirable species such as \ch{LiOH} and \ch{Li2CO3} that passivate the garnet surface \cite{Vema2023_LHX_surface_species,Zhou2024_LHX_CO2,Li2022_garnet_surface_DFT}. This surface contamination increases the interfacial impedance of the battery, ultimately degrading its performance during charge and discharge \cite{Cheng2014_LHX_registance,Wang2024_LHX}.

Although LHX reaction in garnets have been widely observed experimentally \cite{Nyman2010_LHX,Truong2011_tet-first,Gam2014_Li5_LHX,Cheng2018_garnet_recovery}, many fundamental questions remain unresolved. As shown in Figure \ref{fig:LHX_schematic} (b), Li ions in a garnet diffuse through a three-dimensional (3D) network of interconnected tetrahedral (24d) and octahedral (48g/96h) sites that face-share with each other. While some experiments indicate that Li ions in octahedral sites are preferentially exchanged with protons \cite{Cheng2015_LHX_EELS, Li2015_LHX_neutron_diffraction, Hiebl2019_proton_diffusion}, others claim that tetrahedral Li ions are exchanged first \cite{Truong2011_tet-first, Nyman2010_tet-first_IR_NMR, Redhammer2021_tet-first_SCXRD}. Because Li ions in octahedral sites generally have higher site energies \cite{Zeng2022_high_entropy}, the experimental observation of the exchange of tetrahedral Li ions before octahedral ones is counterintuitive. In addition, due to the slow proton diffusion inside the oxide bulk \cite{Ishii2024_H_diffusion, Hiebl2019_proton_diffusion, Gombotz2023_H_diffusion}, garnet samples exposed to water exhibit a gradient of proton concentration from the surface to the bulk \cite{Sharafi2017_depth_profile, Brugge2018_depth}, making accurate measurement of the true thermodynamic equilibrium of the protonated phase experimentally challenging \cite{Ye2021_LHX_review}. In comparison, NASICON-type Li-ion conductors are experimentally shown to be much more stable in an aqueous environment \cite{Safanama2017_LAGP, Ding2012_H_diffusion_in_LATP}, though the origin of this high water stability is also unclear.

In this work, we investigate the thermodynamic origin of LHX reactions for two representative oxide Li-ion conductors, i.e., garnets and NASICONs. We study the site preference of the LHX reaction in the most representative garnet compound \ch{Li7La3Zr2O12}. We further examine the trend observed for \ch{Li7La3Zr2O12} in various other chemical compositions with garnet or NASICON structural frameworks. Leveraging machine-learning interatomic potentials fine-tuned with density functional theory (DFT) data, we extensively sample Li/H/vacancy ordering across compositions with varying Li contents. Our results suggest that Li stuffing significantly increases the driving force of the LHX reaction in garnets, but has less of an effect on the water stability of NASICONs. Our predictions are validated by experiments using inductively coupled plasma mass spectrometry (ICP-MS) measurement on garnets after they have been immersed in an aqueous solution. The higher stability for garnets with lower Li contents suggests a trade-off between high Li-ion conductivity and stability and resistance against \ch{Li+}/\ch{H+} exchange.


\subsection*{Li site preference for \ch{Li+}/\ch{H+} exchange reaction in \ch{Li7La3Zr2O12}}

\begin{figure*}[t]
\centering
\includegraphics[width=0.6\linewidth]{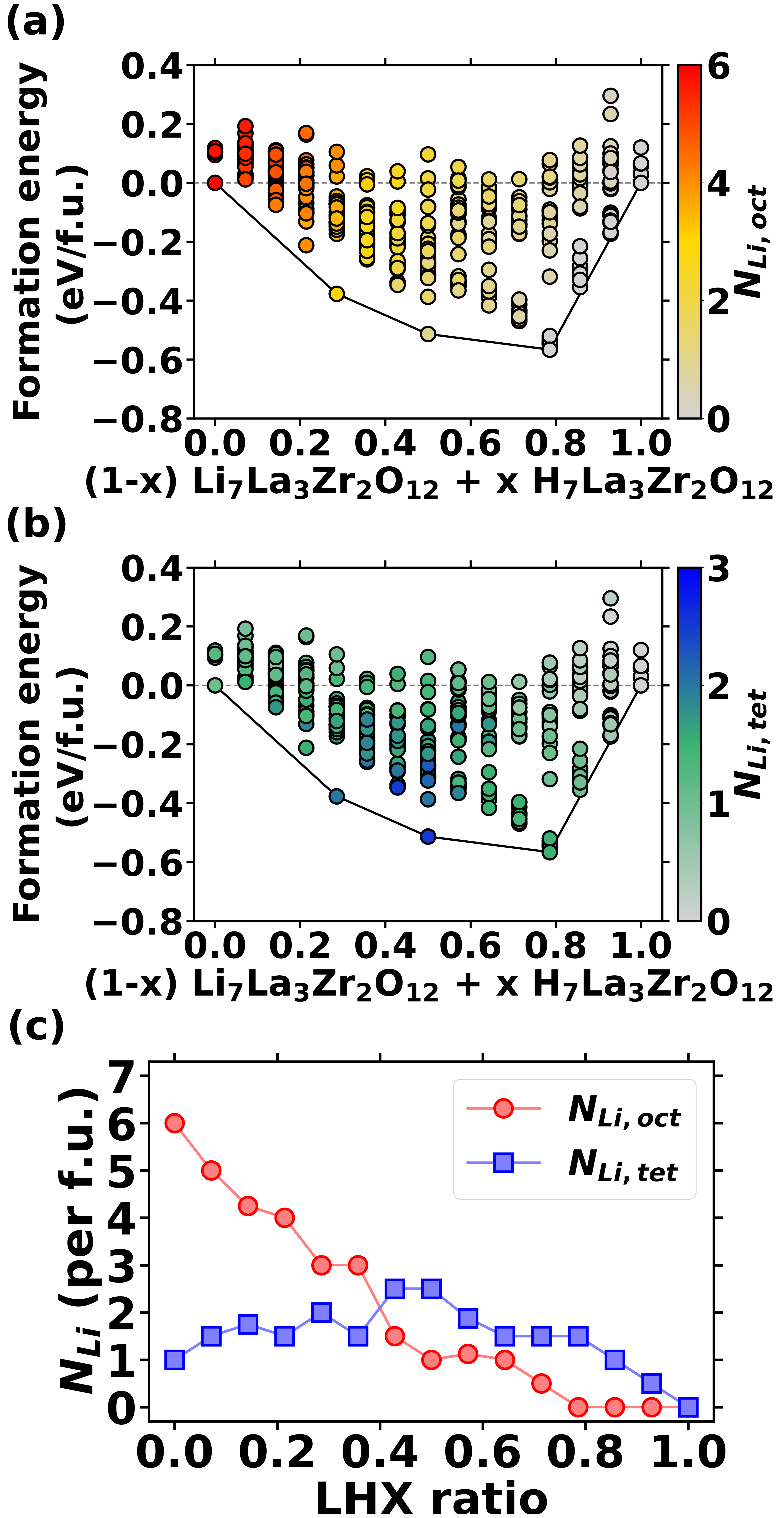}
\caption{\textbf{Li site preference for LHX reaction in \ch{Li7La3Zr2O12}} In (a) and (b), formation energy of partially-exchanged phases (\ch{Li_{7-7x}H_{7x}La3Zr2O12}) are shown as a function of LHX ratio (x) using two different color schemes: (a) The number of remaining Li in  octahedral sites ($N_{Li,oct}$), (b) The number of remaining Li in  tetrahedral sites ($N_{Li,tet}$).  (c) $N_{Li,oct}$ and $N_{Li,tet}$ of the lowest-energy configuration at each LHX ratio x.}
\label{fig:LHX_formation_energy_site_preferance} 
\end{figure*}

In garnets, occupancy of tetrahedral or octahedral sites by Li ions depends on the total Li content per formula unit (f.u.) \cite{Thangadurai2014_garnet_review}. At the lowest Li content (e.g., \ch{Li3La3Te2O12}), Li fully occupies the tetrahedral sites, leaving all octahedral sites vacant. This specific Li/vacancy ordering results in a high activation barrier for a Li-ion to hop from a tetrahedral to a neighboring face-sharing octahedral site, leading to a negligible Li-ion conductivity at room temperature. The Li-ion conductivity quickly increases once additional Li ions are added to the structure, which is typically achieved by doping at the \ch{Te^{6+}} site with lower-valence cations, such as \ch{Ta^{5+}} and \ch{Zr^{4+}}. The highest Li-ion conductivity of $\sim$0.1 mS/cm at room temperature can be obtained for garnets with 6–7 Li ions per f.u. (see Figure S1 in Supporting Information). As schematically illustrated in Figure \ref{fig:LHX_schematic} (b), this additional Li occupies octahedral sites that face-share with a neighboring Li ion in the tetrahedral site. The electrostatic repulsion between Li ions also displaces the other face-sharing tetrahedral Li ion into an empty octahedral site. Experimental data indicates that the Li-ion occupation of octahedral sites increases linearly with total Li content, while the occupation of tetrahedral sites decreases \cite{Thangadurai2014_garnet_review}. Stuffing of the garnet with Li creates local high-energy states—previously referred to as activated local environments \cite{Xiao2021_ALE}. Consequently, Li ions in Li-stuffed garnets diffuse via concerted propagation of these local structural motifs. Such a concerted diffusion exhibits a significantly lower energy barrier compared to a single Li-ion hop in Li-unstuffed garnets \cite{He2017_fast_conduction}.
 
Unlike Li ions, protons are known to covalently bind to a single oxygen forming a hydroxyl group \cite{Kitchaev2017_MnO2_protonation}. Hence one expects them to have less of a preference for a particular oxygen coordination as is the case for Li. The lack of site preference for the proton would imply that the first Li ions to be exchanged should be from the higher energy octahedral sites. Indeed, the evidence for octahedral-first exchange has been provided by various experimental techniques, including electron energy loss spectroscopy (EELS) \cite{Cheng2015_LHX_EELS}, nuclear magnetic resonance (NMR) spectroscopy \cite{Larraz2015_NMR_LLZO}, neutron diffraction \cite{Li2015_LHX_neutron_diffraction}, and single-crystal X-ray diffraction (SCXRD) \cite{Hiebl2019_proton_diffusion}. However, contrasting results have also been reported in both infrared (IR) and NMR spectroscopy studies \cite{Truong2011_tet-first, Nyman2010_tet-first_IR_NMR} and in a more recent SCXRD study \cite{Redhammer2021_tet-first_SCXRD}, where preferential extraction of tetrahedral Li ions was observed. The contrasting experimental results reflect the inherent challenge in detecting light elements, such as Li and H. X-ray techniques are generally less sensitive to these elements than to heavier elements \cite{Ren2018_XRD_ND}. Neutron scattering offers better elemental contrast between neighboring elements in the periodic table; however, the large incoherent scattering of hydrogen decreases the signal-to-noise ratio \cite{Ren2018_XRD_ND,Wang2022_ND}, complicating detection of \ch{Li+}/\ch{H+} exchange. Solid-state NMR and vibrational spectroscopies often produce overlapping features or low signal-to-noise ratio for dilute hydrogen and lithium environments \cite{Nyman2010_tet-first_IR_NMR,Larraz2015_NMR_LLZO}.

Here we perform density functional theory (DFT) calculations to investigate which Li/H/vacancy configuration is thermodynamically more favorable as LHX proceeds. More specifically, DFT calculations are conducted to evaluate zero-K formation energies of partially exchanged garnets \ch{Li_{7-7x}H_{7x}La3Zr2O12} with respect to two endpoint compounds, i.e., the pristine (\ch{Li7La3Zr2O12}) and the fully protonated (\ch{H7La3Zr2O12}) phases. For the partially exchanged structures, we enumerated various Li/H/vacancy configurations where Li ions in octahedral and/or tetrahedral sites are exchanged with varying ratios (see Supporting Information). The calculated formation energies of partially exchanged phases \ch{Li_{7-7x}H_{7x}La3Zr2O12} are presented in Figures \ref{fig:LHX_formation_energy_site_preferance}(a) and (b) with two different color schemes: in Figure \ref{fig:LHX_formation_energy_site_preferance} (a), the color transitions from red to gray as the number of Li in octahedral sites ($N_{\mathrm{Li,oct}}$) decreases, while in Figure \ref{fig:LHX_formation_energy_site_preferance} (b), the color changes from blue to gray as the number of Li in tetrahedral sites ($N_{\mathrm{Li,tet}}$) decreases. Figures \ref{fig:LHX_formation_energy_site_preferance} (a) and (b) show that many partially exchanged structures have negative formation energies, collectively constructing the energy convex hull. This indicates that partial proton exchange is thermodynamically favorable, with the extent of exchange depending on the Li and H chemical potentials of surrounding environments. The values of $N_{\mathrm{Li,oct}}$ and $N_{\mathrm{Li,tet}}$ in the lowest-energy configurations are shown as a function of LHX ratio x in Figure \ref{fig:LHX_formation_energy_site_preferance} (c). Our results indicate that octahedral Li ions are progressively replaced with protons from the early stage of LHX. On the other hand, the number of tetrahedral Li ions slightly increases until $\text{x}\approx0.5$, and then decreases for higher LHX ratios. The initial increase in $N_{Li,tet}$ indicates that some octahedral Li ions migrate into empty tetrahedral sites upon LHX to reduce the energy. This is also evident in Figure \ref{fig:LHX_formation_energy_site_preferance}(b), where lower-energy configurations have higher $N_{\mathrm{Li,tet}}$ values at $\text{x}\approx0.5$. The increase in $N_{Li,tet}$ during the early stage of LHX is consistent with the experimental observation that in pristine, non-exchanged garnets, the number of octahedral (tetrahedral) Li ions decreases (increases) as the total Li content decreases \cite{Thangadurai2014_garnet_review}, which is attributed to the higher Li site energy in octahedral sites \cite{Zeng2022_high_entropy}. Our results imply that the Li site energy is also the dominant factor in determining which Li ions are exchanged first during proton exchange. As shown by the calculated results in Figure S2 in Supporting Information, in a perfect cubic host structures of \ch{La3Zr2O12}, the site energy of an octahedral Li ion is $\sim$0.9 eV higher than that of a tetrahedral Li ion, whereas the site energy of protons are identical across all oxygen anions they bond to. In other words, the energy to exchange a Li ion with a proton in a garnet depends primarily on the site the Li ion occupied, and is less sensitive to the proton site. Based on our results, we conclude that the octahedral-first proton exchange is thermodynamically more favorable.

\FloatBarrier
\subsection*{\ch{Li+}/\ch{H+} exchange energy for oxide Li-ion conductors with varying Li contents}

So far, we have demonstrated that octahedral Li ions are thermodynamically less resistant to proton exchange than tetrahedral Li. Given that the number of octahedral Li in a garnet structure decreases as the total Li content decreases \cite{Thangadurai2014_garnet_review}, we further investigate whether garnet compounds with lower total Li content per f.u. exhibit better water stability. As a matter of comparison, we also evaluate if NASICONs exhibit a similar Li content-dependence on the stability against proton exchange. Similar to garnets, Li ions can be stuffed into the baseline NASICON compound \ch{LiTi2(PO4)3} to increase the Li-ion conductivity \cite{Arbi2013_NASICON_structure,Arbi2014_LATP,Lang2015_LATP,Xiao2021_ALE}, with the highest Li-ion conductivity typically observed at compositions containing $\sim$1.3 Li per f.u. \cite{Rossbach2018_NASICON_review}. The compositions we consider here include \ch{Li3La3Te2O12} (Li3 Garnet), \ch{Li4La3TaTeO12} (Li4 Garnet), \ch{Li5La3Ta2O12} (Li5 Garnet), \ch{Li6La3ZrTaO12} (Li6 Garnet), \ch{Li7La3Zr2O12} (Li7 Garnet), \ch{LiTi2(PO4)3} (Li1 NASICON), and \ch{Li2TiIn(PO4)3} (Li2 NASICON), covering the range of Li contents of both the baseline compositions (Li3 Garnet and Li1 NASICON) and those Li-stuffed compositions associated with the highest ionic conductivities for both garnet (Li content of 6-7 per f.u. \cite{Thangadurai2014_garnet_review}) and NASICON (Li content of 1-2 per f.u. \cite{Rossbach2018_NASICON_review}) frameworks.

In order to quantify the driving force of the LHX reaction, we define the LHX energy ($E_{\text{LHX}}$) as follows
\begin{equation}
E_{\text{LHX}}(c,y) = (E[c-\text{y}\cdot \text{Li} + \text{y}\cdot \text{H}] + \mathrm{y}\cdot\mu_{\mathrm{Li^+}} - E[c] - \mathrm{y}\cdot\mu_{\mathrm{H^+}})/\mathrm{y}
\label{eq:LHX}
\end{equation}
where $E[c-\text{y}\cdot \text{Li} + \text{y}\cdot \text{H}]$ and $E[c]$ are the energies of the solid phases before and after the LHX reaction, $\mathrm{y}$ is the amount of Li ions exchanged, and $\mu_{\mathrm{Li^+}}$ and $\mu_{\mathrm{H^+}}$ are the chemical potentials in solution which depend on the Li concentration and pH, respectively (see Supporting Information) \cite{Persson2012_pbx,Sun2019_pbx}. To calculate LHX energy, we include the lowest-energy pristine and fully protonated structures, and all partially exchanged structures that lie on the energy convex hull.

To sample enough configurations to construct an accurate energy convex hull, we utilize the Crystal Hamiltonian Graph neural Network (CHGNet) model \cite{Deng2023_CHGNet} to perform structural relaxations of a vast range of Li/H/vacancy configurations (see Supporting Information). CHGNet is a universal machine-learning interatomic potential pre-trained on the Materials Project \cite{Jain2013_MP,Horton2025_MP} trajectory dataset. Here, we further fine-tune the CHGNet model on DFT energies of various pristine and protonated garnet and NASICON structures. Leveraging the fast relaxations possible with CHGNet models, up to 1000 Li/H/vacancy configurations are enumerated for each LHX ratio, which is varied with a step size of one Li per supercell that contains 96 oxygen anions for garnets and 72 for NASICONs. An exception is \ch{LiTi2(PO4)3}, which only contains 6 Li ions per supercell, and all Li ions fully occupy the octahedral (6b) sites. For this compound, all symmetrically inequivalent Li/H/vacancy orderings are enumerated and relaxed with DFT. 

\begin{figure*}[t]
\centering
\includegraphics[width=\linewidth]{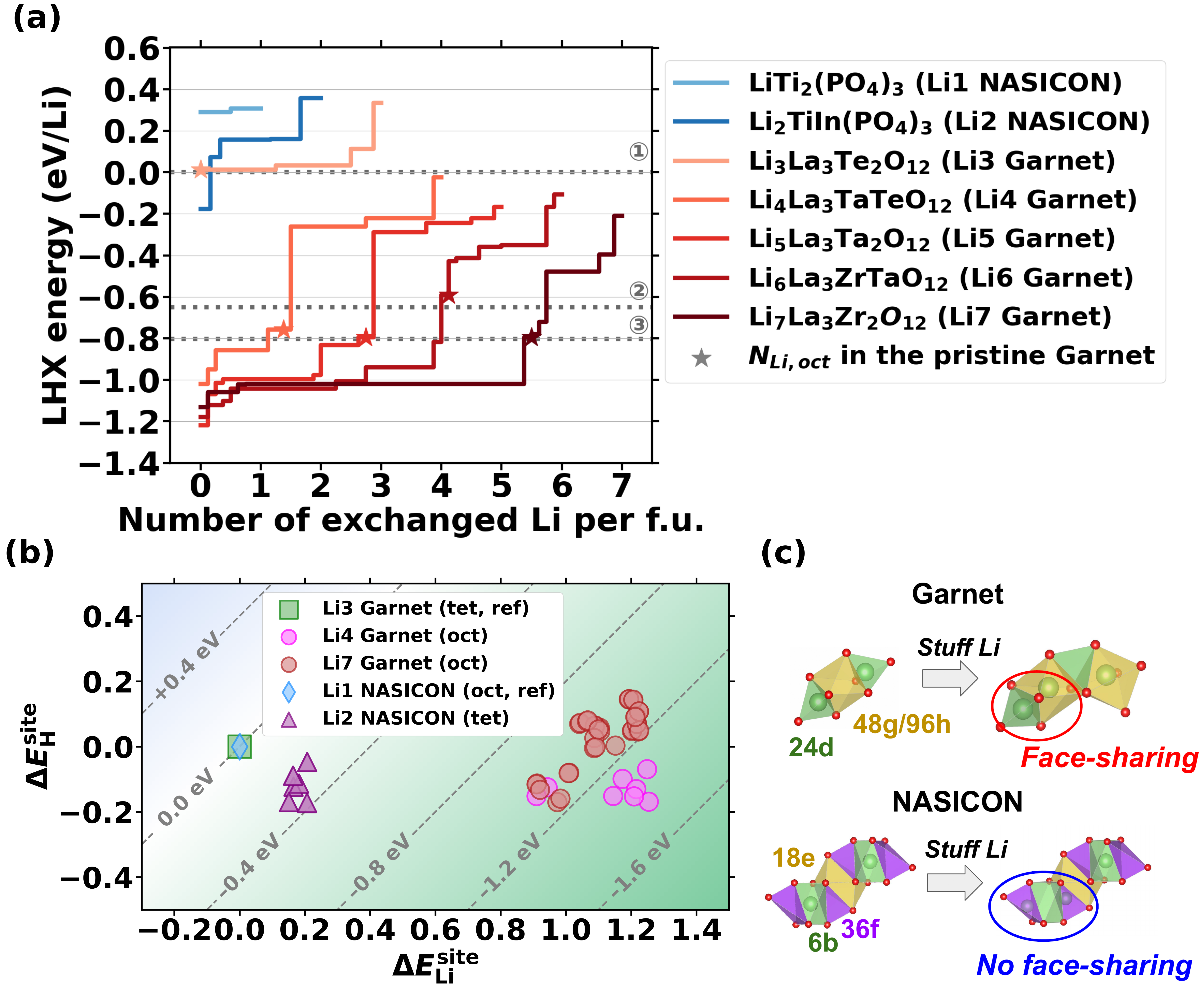}
\caption{\textbf{Effect of Li stuffing on \ch{Li+}/\ch{H+} exchange reaction} (a) LHX energy ($E_{\text{LHX}}$) as a function of total number of exchanged Li per f.u. for various garnet and NASICON compounds with different Li contents. The LHX energy value depends on the chemical potentials $\mu_{\mathrm{Li^+}}$ and $\mu_{\mathrm{H^+}}$, which in turn are function of Li-ion concentration ($C_{\text{Li}^+}$) and pH, respectively. The dashed lines correspond to the reference energies for different conditions: \ding{172} neutral water condition: $\text{C}_{\text{Li}^+}=10^{-6}$ M, pH = 7, \ding{173} strongly alkaline condition: $\text{C}_{\text{Li}^+}=1$  M, pH = 12, and \ding{174} extremely alkaline condition: $\text{C}_{\text{Li}^+}=5$  M, pH = 15. The star represents the number of octahedral Li ($N_{\mathrm{Li,oct}}$) in the pristine, non-exchanged garnet prior to structural relaxation, which closely aligns with the experimental measurements \cite{Thangadurai2014_garnet_review}. (b) Relative Li and H site energies ($\Delta E_{\text{Li}}^{\text{site}}$ and $\Delta E_{\text{H}}^{\text{site}}$) in the Li-stuffed garnet and NASICON structures (Li7 Garnet, Li4 Garnet, or Li2 NASICON) with respect to the values in the unstuffed structures (Li3 Garnet or Li1 NASICON). The blue-to-green color gradient in the background shows the variation of the resulting relative LHX energy of Li-stuffed compounds with respect to the values in the unstuffed structures. Some relative LHX energy values are also labeled with numbers and dashed lines. (c) Li stuffing in a garnet framework leads to the emergence of face-sharing Li–Li pairs between neighboring tetrahedral and octahedral sites. In contrast, a stuffed Li ion in a NASICON framework occupies a tetrahedral site, which displaces the original Li ion from the octahedral site to a neighboring tetrahedral site.}
\label{fig:LHX_energy}
\end{figure*}

Unlike garnets, whose Li occupation in octahedral and tetrahedral sites as a function of total Li content are well characterized \cite{Thangadurai2014_garnet_review}, there is no general consensus on the Li site occupancy among the 6b, 36f, and 18e sites in Li-stuffed NASICONs (schematically illustrated in Figure \ref{fig:LHX_energy}(c)) \cite{Arbi2013_NASICON_structure, Arbi2014_LATP, Lang2015_LATP, Rossbach2018_NASICON_review, Sugantha1997_Li2_NASICON}. For this reason, we used various Li/vacancy ordering enumeration schemes and selected the 20 Li/vacancy configurations that have the lowest DFT-relaxed energy in \ch{Li2TiIn(PO4)3} (see Figure S6 in Supporting Information). Our results indicate that the tetrahedral sites (36f) are preferentially occupied for most of the low energy configurations. In fact, all Li ions occupy 36f sites in the lowest-energy structure (see Table S5 in Supporting Information). LHX energy ($E_{\text{LHX}}$) for \ch{Li2TiIn(PO4)3} is further calculated by partially or completely exchanging Li by protons in these 20 low-energy configurations. Note that previous experimental measurements indicate that the symmetry of NASICON framework in the \ch{Li_{1-x}Ti_{2-x}In_x(PO4)3} series varies as the stuffed Li content x increases \cite{Sugantha1997_Li2_NASICON, Hamdoune1986_LTIP_structure}. For 0 $<$ x $<$ 0.4, the compounds retain the original rhombohedral $R\overline{3}C$ structure. With further Li stuffing, the crystal symmetry changes to orthorhombic (Pbca) for 0.4 $<$ x $<$ 1.0 and to monoclinic (P2$_1$/n) for 1.0 $<$ x $<$ 2.0 \cite{Hamdoune1986_LTIP_structure}. In our calculations, we use the rhombohedral structure for both \ch{LiTi2(PO4)3} and \ch{Li2TiIn(PO4)3}, so that the effect of Li stuffing can be compared without being affected by the difference in crystal symmetry.

Figure \ref{fig:LHX_energy} (a) shows the calculated LHX energy ($E_{\text{LHX}}$) as a function of total number of exchanged Li per f.u. for various garnet and NASICON compounds. In this plot, the LHX energies are referenced to the chemical potentials $\mu_{\mathrm{Li^+}}$ and $\mu_{\mathrm{H^+}}$ in neutral water with $\text{C}_{\text{Li}^+}=10^{-6}$ M and pH = 7 (highlighted with the dashed line \ding{172}). As can be seen, the LHX energy generally increases with increasing number of exchanged Li ions, indicating that the LHX reaction becomes progressively less favorable as more Li ions are exchanged. Nonetheless, most reaction energies are negative for the garnets, indicating that proton exchange with Li in garnets is generally favorable in neutral water. Compounds with a higher Li content tend to exhibit more negative LHX energies for both garnets and NASICONs. Li3 Garnet (the lowest Li-content garnet) exhibits the most positive LHX energies among all garnets, with LHX energies remaining positive across the whole range of number of exchanged Li. However, once additional Li ions are stuffed (Li4--7 garnets), the LHX energy value drops to a negative value, especially during the initial stage of LHX reaction: near $N^{\text{X}}_{\text{Li}}\approx0$, the LHX energy decreases from $\sim$0.0 eV/Li for the Li3 Garnet to less than $-1.0$ eV/Li for all the other Li-stuffed garnets, indicating that the additional Li needed to achieve good conductivity is susceptible to proton exchange. For these Li-stuffed garnets, the LHX energy forms a plateau at $E_{\text{LHX}}$ that ranges from $-1.0$ to $-0.8$ eV/Li depending on the Li content. The plateau ends approximately when all Li ions in the octahedral sites ($N_{\mathrm{Li,oct}}$) are exchanged. Beyond this point, there is a sharp transition of the LHX energies into a second plateau at $E_{\text{LHX}}$ that ranges from $-0.5$ to $-0.2$ eV/Li, which corresponds to the LHX energy for exchanging tetrahedral Li ions in a given garnet. Notably, for all Li-stuffed garnets, the LHX energies remain negative even at the end of the LHX reaction, suggesting that even a complete exchange of Li ions is thermodynamically viable for these Li-stuffed compounds. 

In comparison, the LHX energy values for NASICONs are mostly positive and even higher than that of Li3 Garnet, indicating that NASICONs are in general more stable against proton exchange compared to garnets. Similar to garnets, the baseline NASICON compound (Li1 NASICON) also has a higher LHX energy plateau than the Li-stuffed composition (Li2 NASICON), suggesting that the negative effect of Li-stuffing on the water stability is a general issue among oxide conductors.

The absolute value of $E_{\text{LHX}}$ depends on the chemical potentials of \ch{Li+} and \ch{H+} in solution (see Equation \ref{eq:LHX}), which vary with external conditions of $\text{C}_{\text{Li}^+}$ and pH. Specifically, a higher $\text{C}_{\text{Li}^+}$ increases $\mu_{\mathrm{Li^+}}$, while a higher pH decreases $\mu_{\mathrm{H^+}}$, both of which lead to a shift of $E_{\text{LHX}}$ towards a more positive value, and thus stabilize materials against proton exchange. The reference energies at which the reaction energy is zero are labeled for higher $\text{C}_{\text{Li}^+}$ and pH conditions in Figure \ref{fig:LHX_energy} (a) as \ding{173} for $\text{C}_{\text{Li}^+}=1$ M and pH = 12 (referred to as the strongly alkaline condition), and \ding{174} for $\text{C}_{\text{Li}^+}=5$ M and pH = 15 (referred to as the extremely alkaline condition). The pH in the strongly alkaline condition is the value that can be attained for a \ch{LiOH} solution buffered by \ch{H3BO3} \cite{Lam2024_degratation_LTGP}. The even higher pH value in the extremely alkaline conditions is the maximum limit that can be achieved by a saturated LiOH solution, above which the solid phase of LiOH precipitate out at room temperature \cite{Monnin2005_LiOH_PD}. As shown in Figure \ref{fig:LHX_energy}(a), both of these highly alkaline conditions shift down the reference energy by more than 0.6 eV/Li compared to the neutral pH condition. These extreme conditions can effectively protect Li ions in tetrahedral sites from being exchanged; however, the octahedral Li ions remain susceptible to the LHX reaction due to their even more negative $E_{\text{LHX}}$ values. These results suggest that the driving force for the LHX reaction in garnets is so strong that even extremely high levels of $\text{C}_{\text{Li}^+}$ and pH cannot completely suppress it. In fact, many garnet materials gradually undergo proton exchange even in ambient or humid air without direct exposure to liquid phase water. If gas phase \ch{H2O} and solid LiOH are considered as the reactant and product of the proton exchange reaction, respectively, the reference energy at which the reaction energy is zero shifts down by about –0.9 eV relative to the neutral water condition \ding{172} (see Supporting Information). This energy shift is close to that under extremely alkaline conditions \ding{174} (–0.8 eV) and remains slightly more positive than the LHX energy plateau for the octahedral Li ions in Li-stuffed garnets (–1.0 eV). Therefore, the susceptibility of octahedral Li ions to atmospheric humidity can be considered to be on par with their behavior in highly alkaline solutions.

For a given $\text{C}_{\text{Li}^+}$ and pH condition, the LHX energy solely depends on the difference of the two energy terms $E[c-\text{y}\cdot \text{Li} + \text{y}\cdot \text{H}]$ and $E[c]$ (see Equation \ref{eq:LHX}). Therefore, the LHX energy essentially captures the energy difference between incorporating H versus Li into a given structure. To better understand the difference in the LHX driving force between garnets and NASICONs, we perform DFT calculations to compare the Li and H site energies ($E_{\text{Li}}^{\text{site}}$ and $E_{\text{H}}^{\text{site}}$), which is estimated as the negative of energy required to extracting a single Li or H from a given site in a stoichiometric structure (see Supporting Information). This site energy varies across different Li sites, reflecting differences in intrinsic site energies between octahedral and tetrahedral sites, as well as interactions between the inserted species and neighboring Li ions. Also, because each Li site is coordinated by multiple oxygen anions (four for tetrahedral and six for octahedral), we start the structural optimization with H in a position close to each oxygen, from which we obtain a range of H site energy for each Li site (see Figure S5 in Supporting Information). We report the lowest energy value as the H site energy of a given Li site, assuming that there is enough time for the system to reach the thermodynamic equilibrium during proton exchange. Due to the high symmetry, the H site energy converges to a single value in the unstuffed Li3 Garnet and Li1 NASICON compounds.

In Figure \ref{fig:LHX_energy} (b), the calculated Li and H site energies at the high-energy site of the Li-stuffed compounds (octahedral sites of Li7 Garnet, octahedral sites of Li4 Garnet, or tetrahedral sites of Li2 NASICON) are shown. Each data point represents a specific Li site in a given structure. The zero of energies of Li and H are selected as the site energies in the corresponding unstuffed baseline compounds (tetrahedral sites of Li3 Garnet or octahedral sites of Li1 NASICON), so that relative Li and H site energies between the Li-stuffed and unstuffed compositions ($\Delta E_{\text{Li}}^{\text{site}}$ and $\Delta E_{\text{H}}^{\text{site}}$) are compared. The LHX energy of Li-stuffed compounds relative to the unstuffed ones, determined by the difference between relative Li and H site energies for each site ($\Delta E_{\text{H}}^{\text{site}}-\Delta E_{\text{Li}}^{\text{site}}$), are also indicated by the gradient of color in the background and the diagonal dashed lines. The figure clearly illustrates how Li-stuffing in garnet raises the chemical potential of Li ions. The Li site energy can be more than 1.0 eV higher in the Li-stuffed garnet framework compared to the unstuffed one. Both Li4 and Li7 Garnets exhibit a similar magnitude of increase in the Li site energy, implying that the Li site energy is mostly determined by the local face-sharing Li-Li configuration and is less sensitive to the total number of stuffed Li ions. In contrast, figure \ref{fig:LHX_energy} (b) shows that the H site energy does not vary as much upon Li stuffing. The Li-stuffed structures have many sites for which the H site energy is even slightly lower (down to $\sim -0.2$ eV) than in the corresponding unstuffed structures. Although not explicitly calculated, considering the similar Li site energies  (ranging from 0.9 to 1.2 eV) and H site energies (ranging from –0.2 to 0.2 eV) calculated for the least stuffed (Li4 Garnet) and most stuffed (Li7 Garnet) garnet compositions, we expect that the Li and H site energies in the intermediate compositions (Li5 Garnet and Li6 Garnet) should also fall within these ranges

In contrast to the garnets, the increase in Li site energy by Li stuffing is only $\sim$0.2 eV for the NASICON. We attribute this difference between the two crystal frameworks to the distinct local Li-ion configurations in the Li-stuffed structures \cite{Xiao2021_ALE, Lang2015_LATP}. As shown in Figure \ref{fig:LHX_energy}(c), Li stuffing in the garnet framework leads to face-sharing Li–Li pairs between neighboring tetrahedral (24d) and octahedral (48g/96h) sites. Our calculations show that the Li-Li face-sharing energy is at least $\sim$0.45 eV per face-sharing pair (see Figure S4 in Supporting Information), which contributes to the Li site energy in addition to the intrinsic site energy difference between octahedral and tetrahedral sites in garnets. In contrast, in the NASICON framework, Li stuffing displaces the original Li ion from the octahedral site (6b) into an adjacent tetrahedral site (36f), thereby avoiding direct face-sharing between the original and stuffed Li ions. In other words, NASICONs have much sparser Li-ion configurations with larger Li-Li distance compared to garnets, resulting in smaller electrostatic interaction between Li-ions, effectively lowering the Li site energies. This is also reflected in their composition, as NASICONs accommodate much fewer Li ions per f.u. compared to garnets. 

\FloatBarrier
\subsection*{Effect of chemical substitutions}

\begin{figure*}[t]
\centering
\includegraphics[width=\linewidth]{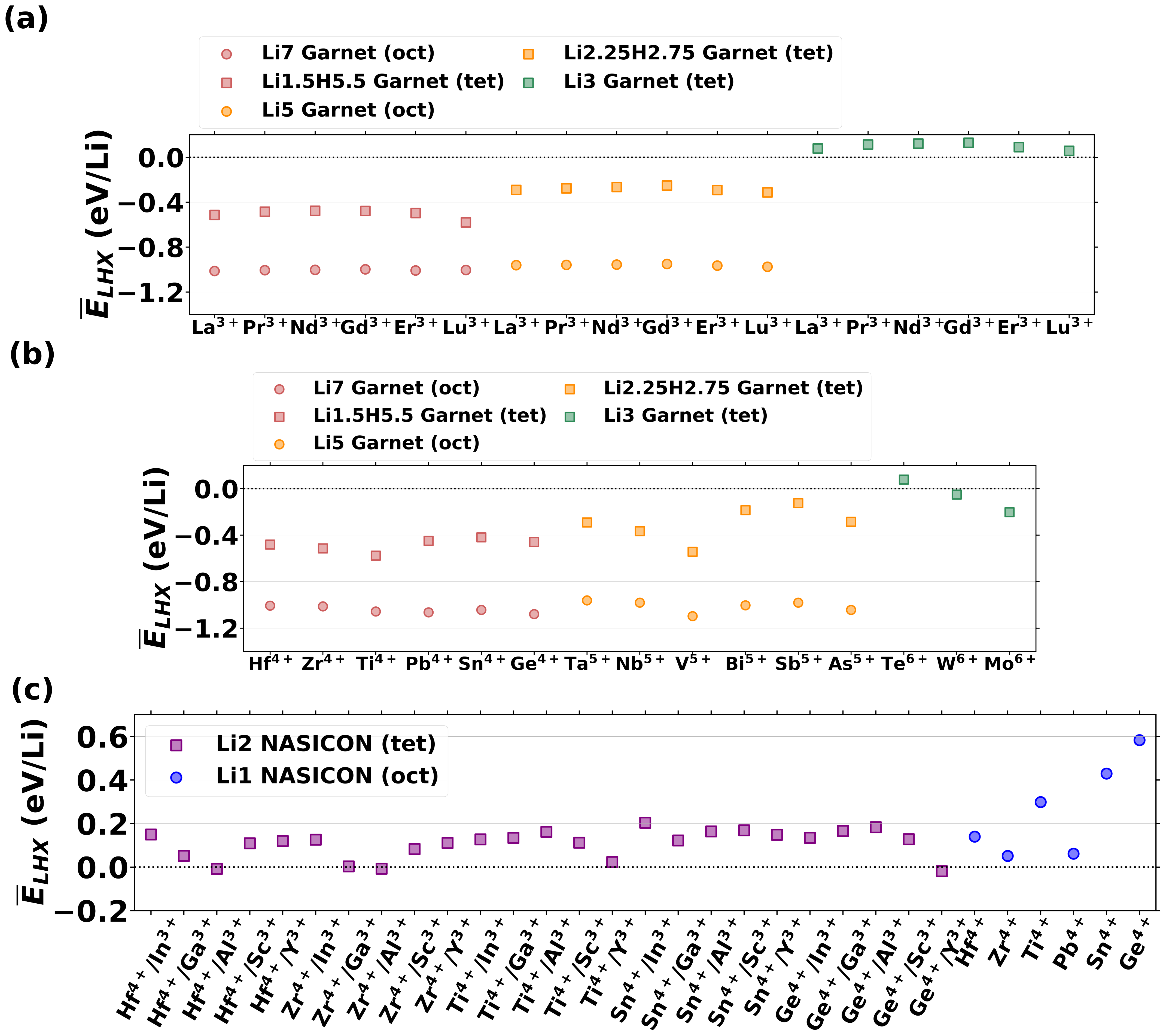}
\caption{\textbf{Effect of chemical substitution} The average LHX energy ($\overline{E}_{\text{LHX}}$) is calculated at different type of Li sites for various cation substitutions. Substitutions at (a) La site and (b) Zr site of garnet, and (c) Ti site of NASICON. For Li2 NASICONs, we perform double substitutions, and the two substituted cations are labeled using a slash. The chemical potentials for Li and H are set at the neutral water condition ($\text{C}_{\text{Li}^+}=10^{-6}$ M and pH=7)}
\label{fig:substitution} 
\end{figure*}

We investigate whether chemical substitutions of the non-Li cations in both garnets and NASICONs can modify the LHX energy in a substantial way. In Figure \ref{fig:substitution} (a), we show the calculated average LHX energy ($\overline{E}_{\text{LHX}}$) for garnets with different $+3$ rare earth elements substituted for \ch{La3+}. Similar results are shown in Figure \ref{fig:substitution} (b) for substitution of the other cations (i.e., \ch{Zr^{4+}}, \ch{Ta^{5+}}, or \ch{Te^{6+}}) in the garnet. For garnets, the $\overline{E}_{\text{LHX}}$ value is evaluated separately for octahedral and tetrahedral Li ions. The exchange energy for the octahedral sites is obtained by replacing all octahedral Li ions with protons, while the value for the tetrahedral sites is determined by replacing the remaining tetrahedral Li ions with protons. For each pristine or proton-exchanged structure, we used the same Li/H/vacancy configuration in the structure with the lowest CHGNet energy obtained from the calculations shown in Figure \ref{fig:LHX_energy}. The chemical substitution is performed by replacing the original non-Li cations without changing the Li/H/vacancy configuration, followed by DFT relaxation. Our results show that chemical substitution has only a minor impact on the average LHX energy. Consistent with the results shown in Figure \ref{fig:LHX_energy}, the $\overline{E}_{\text{LHX}}$ value for tetrahedral Li ions is higher than that of octahedral Li ions for all cation substitutions. However, all $\overline{E}_{\text{LHX}}$ values for both octahedral and tetrahedral sites are negative, indicating that all Li ions in Li-stuffed garnets are potentially susceptible to proton exchange irrespective of the cation chemistry. In contrast, the $\overline{E}_{\text{LHX}}$ values for the unstuffed garnet (Li3 Garnet) remain positive, except for \ch{W^{6+}} and \ch{Mo^{6+}} substitutions which exhibit slightly negative values. 

The results for the average LHX energy when different cations are substituted for \ch{Ti^{4+}} and/or \ch{In^{3+}} in the NASICON structures are shown in \ref{fig:substitution} (c). Since in NASICONs, all Li ions occupy octahedral sites in the Li1 NASICON and tetrahedral sites in the lowest-energy Li2 NASICON (see Table S5), we use the pristine and the fully exchanged structures to calculate $\overline{E}_{\text{LHX}}$. Similar to the garnets, chemical substitution in the NASICONs also does not result in a significant change of $\overline{E}_{\text{LHX}}$. Most of NASICON compounds exhibit a positive $\overline{E}_{\text{LHX}}$ value, suggesting that NASICONs are generally stable against proton exchange. Remarkably, \ch{Sn^{4+}} and \ch{Ge^{4+}} substitutions in the Li1 NASICON lead to higher $\overline{E}_{\text{LHX}}$ values than other compositions. The high water stability of \ch{Li_{1-x}Al_xGe_{2-x}(PO4)3} LAGP is indeed previously experimentally demonstrated, with no LHX reaction was observed after 100 h of immersion in water \cite{Safanama2017_LAGP}.

\FloatBarrier
\subsection*{Experimental study of \ch{Li+}/\ch{H+} exchange reactions in garnet compounds}

\begin{figure*}[h]
\centering
\includegraphics[width=\linewidth]{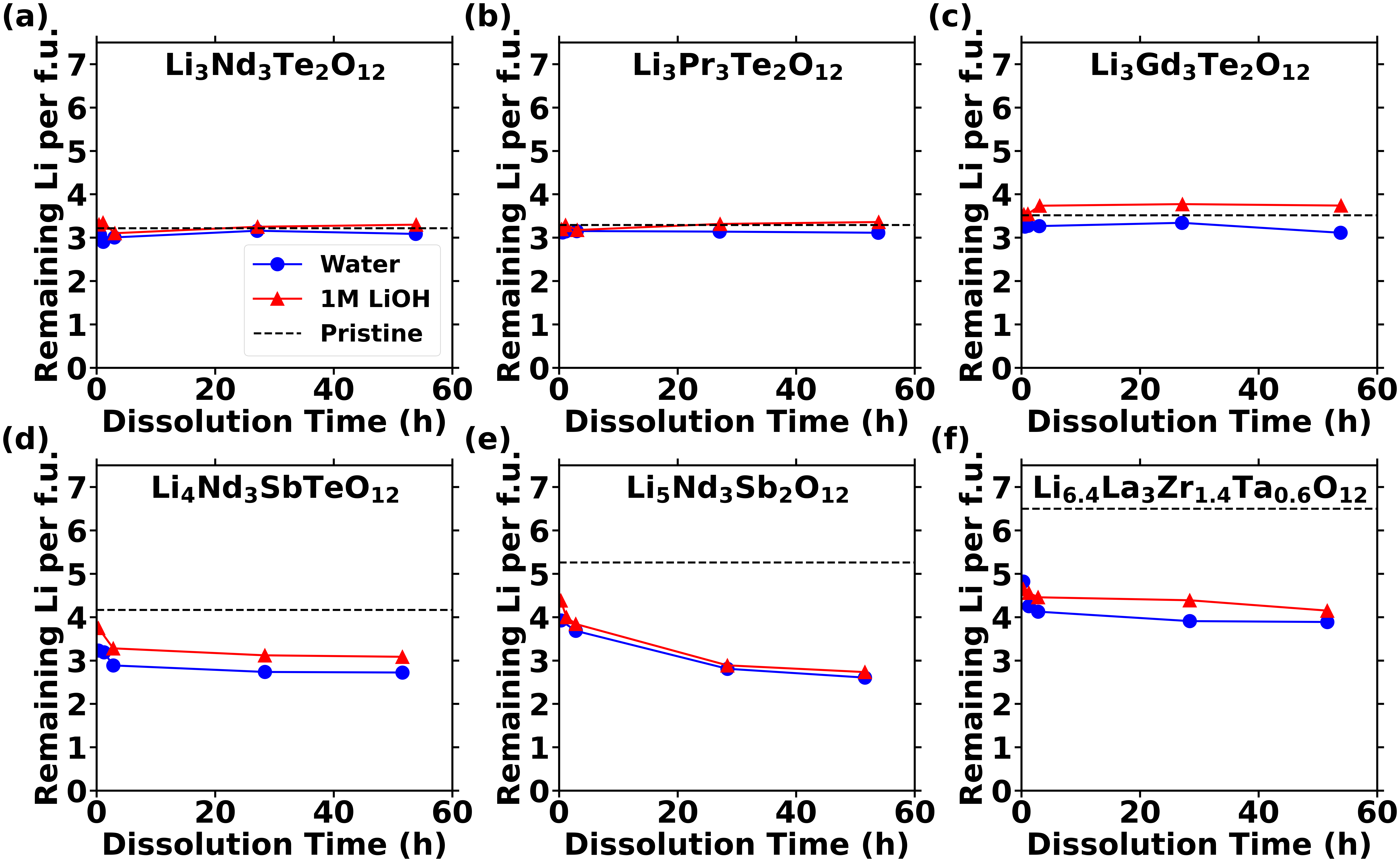}
\caption{\textbf{Experimental measurements of \ch{Li+}/\ch{H+} exchange reactions for garnet compounds} The time evolution of the remaining number of Li content per f.u. of (a) \ch{Li3Nd3Te2O12}, (b) \ch{Li3Pr3Te2O12}, (c) \ch{Li3Gd3Te2O12}), (d)\ch{Li4Nd3SbTeO12}, (e) \ch{Li5Nd3Sb2O12}, and (f)\ch{Li_{6.4}La3Zr_{1.4}Ta_{0.6}O12}. Powder samples of these compositions are synthesized with solid-state synthesis. Samples are immersed in either water or 1M LiOH solution. The remaining number of Li per f.u. in the bulk is measured by ICP-MS.}
\label{fig:exp_Garnet_LHX} 
\end{figure*}

Our calculations predict that Li-stuffed garnets will undergo \ch{Li+}/\ch{H+} exchange in both neutral and alkaline aqueous environments, regardless of the specific cation chemistries. To validate these predictions, we experimentally measure the evolution of Li content by immersing various garnet compounds synthesized via solid-state synthesis in solutions. Specifically, the synthesized powder samples are immersed in either neutral water or a 1 M LiOH solution. The amount of Li remaining in the garnet compounds for various immersion times is measured by inductively coupled plasma mass spectroscopy (ICP-MS). In Figure \ref{fig:exp_Garnet_LHX}, we show these results for both unstuffed (\ch{Li3Nd3Te2O12}, \ch{Li3Pr3Te2O12}, and \ch{Li3Gd3Te2O12}) and Li-stuffed (\ch{Li4Nd3SbTeO12}, \ch{Li5Nd3Sb2O12}, and \ch{Li_{6.4}La3Zr_{1.4}Ta_{0.6}O12}) compositions. The data in Figures \ref{fig:exp_Garnet_LHX} (a-c) show that all unstuffed garnets are stable in water, with almost no detectable loss of bulk Li content after 54 hours of immersion in either neutral or alkaline solutions. In contrast, Figures \ref{fig:exp_Garnet_LHX} (d-f) show that Li-stuffed garnets release Li ions into the solution very quickly, and the remaining Li content gradually levels off with immersion time. Surprisingly, similar results are observed for the immersion in both water and LiOH solution, with only slightly higher remaining Li content when immersed in 1 M LiOH solution. By measuring the pH of the solution after immersing Li-stuffed garnets, we find that both solutions actually equilibrate at a similar pH value of 12–13, with the LiOH solution exhibiting a slightly higher pH (see Figure S10 in Supporting Information). The increase of pH when garnets are exposed to neutral water has been widely observed in experiments \cite{Truong2011_tet-first,Yow2016_pH_increase, Safanama2017_LAGP}, and is due to the uptake of protons into the garnet, leaving \ch{OH-} in the solution. The resulting pH conditions are actually close to the condition \ding{173} ($\text{C}_{\text{Li}^+}=1$  M and pH = 12) used in our calculations shown in Figure \ref{fig:LHX_energy} (a). Under these alkaline conditions, our calculation predicts that all octahedral Li ions should be exchanged with protons, while the tetrahedral Li ions ($\sim$2--3 Li per f.u. for Li4 and Li5 Garnets) remain in the bulk. Experimentally, the measured remaining Li contents for Li-stuffed garnets are 3.1 and 2.7 Li per f.u. for \ch{Li4Nd3SbTeO12} and \ch{Li5Nd3Sb2O12}, respectively, which are close to the number of tetrahedral Li ions in pristine, non-excahnged garnets. A larger discrepancy is observed for the garnets with a higher Li content (\ch{Li_{6.4}La3Zr_{1.4}Ta_{0.6}O12}), which has 4.1 remaining Li per f.u. after the immersion, while the number of tetrahedral Li ions are only $\sim$1--2 Li per f.u. for Li6 and Li7 Garnets. The larger amount of remaining Li may in part originate from the spontaneous migration of octahedral Li to tetrahedral sites upon proton exchange, as was demonstrated in Figure \ref{fig:LHX_formation_energy_site_preferance} (c). Due to the octahedral-to-tetrahedral migration, more Li ions can be stabilized than the original number of tetrahedral Li in the pristine garnets. However, since there are only 3 tetrahedral sites per f.u. in total, the increase in the number of tetrahedral Li alone cannot fully explain the even higher number of remaining Li. This may also be attributed to the slow kinetics of the protons diffusion inside garnets \cite{Hiebl2019_proton_diffusion,Gombotz2023_H_diffusion,Ishii2024_H_diffusion}, preventing the system from reaching the true thermodynamic equilibrium within the experimental timescale. Overall, our experimental results qualitatively agree with our computational predictions, indicating that garnets are generally unstable in water when additional Li is incorporated beyond the baseline composition.

\FloatBarrier


Several key results emerge from our computational study. By directly evaluating the energy to exchange Li ions by protons from a solution, we find that garnets are highly sensitive to \ch{Li+}/\ch{H+} exchange as consistent with observations in the literature. Only the unstuffed, non-conducting garnet with 3 Li per f.u., is predicted to be stable against proton exchange. Any increase in the Li content of the garnet leads to a high enough Li chemical potential to become favorable for proton exchange. The octahedral Li ions which have a higher site energy than the tetrahedral sites in garnets show a particularly high driving force for the exchange. We further demonstrate that the strong proton exchange driving force cannot be mitigated by simple chemical substitutions, implying that the chemical nature of the garnet framework is inherently moisture-sensitive. Our results in Figure \ref{fig:LHX_energy} indicate that this sensitivity is mostly due to the strong Li-Li interaction, which increases the Li chemical potential when the Li content increases. These results also point at the compromise between conductivity and water stability. Li-stuffing enhances Li-ion conductivity by increasing the average energy of Li ions, making the charge carriers more active, but simultaneously destabilizes the material in the presence of water. Such competition poses a fundamental challenge in designing garnet-type Li-ion conductors, which rely on the formation of the ’activated local environment’ of face-sharing Li–Li pairs introduced through Li stuffing to enhance the Li-ion conductivity \cite{Xiao2021_ALE}. The strong increase in Li chemical potential with increasing Li content contrasts with the proton chemical potential, which is mostly unchanged with total proton or Li contents, reflecting the very different bonding of these two species. While \ch{Li+} interacts electrostatically with the host structure and with other Li ions, protons bind strongly to oxygen, making localized covalent bonds \cite{Kitchaev2017_MnO2_protonation}, which is somewhat independent of the environment.

Our calculations also predict that despite the higher Li chemical potential and lower proton chemical potential in a high-pH solution, a highly alkaline environment cannot protect the higher-energy octahedral Li ions in garnets from being exchanged, even though it may offer some protection for the more stable tetrahedral Li ions. The sensitivity of garnet to LHX is further verified by our solution experiments where LHX reaction is observed for all Li-stuffed garnets when immersed in a high-pH LiOH solution. A similar high-pH strategy was previously demonstrated to be effective in preventing LHX reactions for the layered oxide \ch{LiNi_xCo_yMn_{1-x-y}O2} (NCM) \cite{Xu2024_LHX_layered}. In the work by Xu, et al., the LHX reaction of NCM materials could be reversed when the cathode materials were treated in 4M LiOH solutions with high pH and Li-ion concentration. In these NCM crystals, Li ions occupy octahedral sites without face-sharing with each other, and thus the chemical potential can be lower compared to those stuffed Li ions in garnets. In another study, it was shown that protonation of \ch{Li7La3Zr2O12} can be prevented when immersed in hexane solvent whose pKa is as high as 50 \cite{Grissa2021_solvent_impact_LLZO}. This indicates that using an extremely weak organic acid can lower the H chemical potential below that in a saturated LiOH solution, such that even the high-energy Li ions at octahedral sites of garnets can be protected from being exchanged.

Our results show that, in contrast to garnets, NASICON-type Li-ion conductors are expected to be much more resistant to proton exchange as they have either mildly negative or positive LHX energy. We attribute this to several factors. While an absolute site energy for Li cannot easily be defined in materials, we believe that in general Li is more strongly bound in the NASICONs. This is supported by the higher extraction voltage observed in phosphate NASICONs than in simple oxides when the same redox couple is considered \cite{Hautier2011_phosphate_redox}. Furthermore, unlike garnets, NASICONs have no face-sharing Li-Li pairs which increases the energy of Li \cite{Chen2023_spinel,Zhang2020_LTO}. The proton binding energy almost certainly also contributes to the different LHX energetics between garnets and NASICONs. The covalent O-H bond with oxygen is strongly influenced by the other ions that covalently interact with that oxygen, which has been referred to as the inductive effect in the battery literature \cite{Manthiram1989_inductive,Seo2016_O_redox}. In phosphate-based NASICONs, oxygen is strongly covalently bonded to the center P cation, leaving little covalency to make the O-H bond, resulting in weaker O-H bonds. This is unlike the chemistry of garnets in which the oxygen ions mostly interact electrostatically with the other cations. Nonetheless, while proton exchange may not be a problem for phosphate NASICONs, dissolution of \ch{(PO4)3-} is known to be a potential degradation mechanism when in contact with alkaline solution \cite{Lam2024_degratation_LTGP, Mishra2025_LATP_degradation}.

Our findings on lithium NASICONs also shed light on why sodium NASICONs, such as \ch{Na3Zr2Si2PO12}, are generally more susceptible to proton exchange, especially at low pH conditions \cite{Hou2020_sodium_NASICON_water, Bruce1990_sodium_NASICON}. Their higher water instability may stem from the higher alkali metal ion content in sodium NASICONs ($>3$ Na per f.u.) than in the Li-stuffed NASICONs ($1-2$ Li per f.u.), which can potentially raise the average Na energy more significantly. Also, the inductive effect from \ch{(SiO4)^{4-}} is weaker than \ch{(PO4)^{3-}} due to the lower electronegativity of Si compared to P, reducing the Na extraction voltage and enhancing covalent bonding between protons and silicate groups \cite{Masquelier2013_polyanion}. These comparisons point out how subtle differences in crystal frameworks or compositions can strongly influence the water stability of oxide conductors.

Our calculation also shows that the H site energy varies to a certain degree within each polyhedron, depending on which oxygen the proton bonds to (see Figure S5 in Supporting Information). Although the minimum H site energy in a given polyhedron, which sets the driving force for proton exchange, does not vary much with cation chemistry or total Li content as indicated by the results shown in Figure \ref{fig:LHX_energy}, the data in Figure S5 shows that the H site energy can vary by up to 0.7 eV and 0.5 eV for garnet and NASICON, respectively. The variation in H site energy suggests that a proton must overcome energy differences between neighboring O–H bonds, in addition to O-H bond breaking energy, when it diffuses within these oxide conductors. This may partially account for the slow proton diffusion kinetics observed experimentally in protonated garnets (see Figure S1 in Supporting Information). It further implies that when Li ions in the surface region of a garnet are exchanged with protons, the protonated surface layer may become ionically insulating (for both Li ions and protons). While this surface passivation layer may limit further protonation deep within the bulk, it inevitably increases the interfacial resistance and thus negatively impacts battery performance \cite{Brugge2018_depth,Hiebl2019_proton_diffusion,Wang2024_LHX}.

Our results indicate that the potential trade-off between optimizing conductivity and water stability depends on the nature of the conductivity-enhancing mechanism \cite{Xiao2021_ALE, Zeng2022_high_entropy, Jun2024_diffusion_review}. Given our findings, we expect that most simple oxides that derive their high conductivity from “Li-stuffing” will be susceptible to proton exchange in moisture-containing environments. While Li-stuffing can lower the ionic hopping barrier and thereby increase conductivity with little impact on the overall thermodynamic stability of bulk materials \cite{Jun2024_diffusion_review}, our study shows that the increase in the site energy upon Li stuffing can be detrimental when topotactic ionic exchange reactions come into play. These opposing effects call for greater caution in materials design, emphasizing the need to balance improved functionality with long-term chemical stability.

In this light, conductors that rely on other mechanisms, such as the corner-sharing concept \cite{Jun2022_corner_sharing} do not require a large Li content for high conductivity, and may therefore be less water sensitive. The high-entropy effect identified in a variety of structures and chemistries \cite{Zeng2022_high_entropy} is another example for increasing conductivity without increasing Li content, thereby achieving both high conductivity and water stability. In a previous computational high-throughput screening work, some potential superionic oxide Li-ion conductors that possess similar structural characteristics as garnets or NASICONs were identified \cite{Xiao2021_ALE}. Based on our current study, we expect those NASICON-like conductors to exhibit higher water stability than the garnet-like conductors, due to the more homogeneous distribution of Li ions within the diffusion channel that avoids the strong Li-Li interaction.


In summary, we conducted computations combining DFT with a machine-learning interatomic potential model (CHGNet) to investigate the driving force for the \ch{Li+}/\ch{H+} exchange reaction in garnets and NASICONs. We find that Li-stuffed oxides generally exhibit a strong driving force for proton exchange, originating from their high Li chemical potential. In contrast, materials such as NASICONs, which do not possess the face-sharing Li-Li configurations that raises the Li chemical potential and has weaker O-H covalency due to the polyanion-based crystal framework, are predicted to be much less sensitive to proton exchange in humid environments. These findings highlight a fundamental trade-off in the Li-stuffing strategy, where improved Li-ion conductivity may come at the expense of water stability. This work underscores the importance of taking material stability into consideration, rather than focusing solely on optimizing ionic conductivity, when designing materials for practical applications.

\FloatBarrier


\section*{Supporting Information}
Density functional theory methods, Previously reported Li-ion conductivities of various garnet compositions, Li/H site energies in Li-stuffed and unstuffed garnet and NASICON structures, Li and H site energies in cubic garnet frameworks, H site energies in a perturbed host structures of
\ch{La3Te2O12}, Li-Li face-sharing energy, Distribution of H site energy in Li-stuffed Garnet and NASICON, Energy and site occupancy of Li-stuffed Garnets and \ch{Li2TiIn(PO4)3}, Enumeration of Li/H/vacancy configurations, CHGNet model fine-tuning, Li and H chemical potentials in humid air, Density of states, Synthesis and immersion experiments of garnet materials in solutions, Time evolution of solution pH during the immersion of garnet compounds, Calibrations for ICP-MS measurements.

\section*{Author Contributions}
Z.L conducted the computational work. B.X.L. conducted the experiments. S.W. helped analyze experimental data. The manuscript was written by Z.L., and revised by B.X.L., S.W. ,and G.C. The work was supervised by G.C.


\section*{Acknowledgment}
This work was supported by the Samsung Advanced Institute of Technology (SAIT). The computational analysis was performed using computational resources sponsored by the Department of Energy’s Office of Energy Efficiency and Renewable Energy located at the National Renewable Energy Laboratory (NREL). Computational resources were also provided by the Advanced Cyberinfrastructure Coordination Ecosystem: Services \& Support (ACCESS) program, which is supported by National Science Foundation grants \#2138259, \#2138286, \#2138307, \#2137603, and \#2138296.

\setstretch{1}
\bibliography{references}

\end{document}